\documentclass[letterpaper, 10 pt, conference]{ieeeconf}  

\IEEEoverridecommandlockouts                              

\title{\LARGE \bf
On observer forms for hyperbolic PDEs with boundary dynamics}

\author{Luca Mayer and Frank Woittennek%
\\%
{\small Institute of Automation and Control Engineering,
UMIT TIROL -- Private University for Health Sciences and Health Technology}%
\\%
{\small Eduard Walln{\"o}fer Zentrum~1, Hall in Tirol, Austria}%
\\%
{\small \texttt{\{luca.mayer, frank.woittennek\}@umit-tirol.at}}%
\thanks{This research was funded in whole or in part by the Austrian Science Fund (FWF) [10.55776/I6519].}%
}

\usepackage{xspace}
\usepackage{amsfonts}
\usepackage{amssymb}
\usepackage[acronym, nowarn]{glossaries}
\usepackage{todonotes}
\usepackage{bm}
\usepackage{booktabs}
\usepackage{colortbl}
\usepackage{xcolor}
\usepackage{adjustbox}
\usepackage{tikz}
\usepackage{pgfplots}

\usetikzlibrary{plotmarks,math,positioning,shapes,arrows,backgrounds,circuits.logic.IEC,circuits.ee.IEC,decorations.pathmorphing,patterns,shapes.misc,shapes.geometric,calc,fit,spy,matrix,tikzmark,arrows.meta,fillbetween, 3d,decorations.markings}

\pgfplotsset{compat=1.16}

\usepackage{cite}

\DeclareMathSymbol{\shortminus}{\mathbin}{AMSa}{"39}

\newtheorem{proposition}{Proposition}

\newcommand{\pderi}[1][non]{\ensuremath{\ifthenelse{\equal{#1}{non}}{\partial}}{\partial_{#1}}}

\newcommand{\derifr}[2]{\ensuremath{\tfrac{{\rm d}#1}{{\rm d}#2}}}

\newcommand{\dd}[1]{{\, \rm d}#1}

\newcommand{\honftransformation}{\mathcal{T}_{\bar y}^{\eta}}
\newcommand{\invhonftransformation}{\mathcal{T}_{\eta}^{\bar y}}
\newcommand{\timeshiftoperator}[1]{\ensuremath{\rho_{{#1}}}}

\newcommand{\Lin}{\ensuremath{\mathcal{L}}}
\newcommand{\observabilityMap}{\mathcal{T}_{x}^{\bar y}}
\newcommand{\invobservabilityMap}{\mathcal{T}_{\bar y}^{x}}

\newcommand{\reals}{\ensuremath{\mathbb R}}
\newcommand{\naturals}{\ensuremath{\mathbb N}}

\newcommand{\statespace}{\ensuremath{\mathcal H}}
\newcommand{\pdestatespace}{\ensuremath{\mathcal X}}
\newcommand{\odestatespace}{\ensuremath{\reals^n}}

\newcommand{\odestate}{\ensuremath{\bm{\xi}}}
\newcommand{\odeobservabilitystate}{\ensuremath{\bar\odestate}}

\newcommand{\pdestate}{\ensuremath{\bm{x}}}
\newcommand{\transstatefwd}{x^-}
\newcommand{\transstatebwd}{x^+}
\newcommand{\btransstatefwd}{\bar{x}^-}

\newcommand{\sysop}{\mathcal{A}}
\newcommand{\inpop}{\mathcal{B}}
\newcommand{\outop}{\mathcal{C}}
\newcommand{\semigroup}{S}
\newcommand{\reachabilitymap}{R}
\newcommand{\domain}{D}

\newcommand{\systemstate}{\ensuremath{\bm X}}

\newcommand{\transtime}{\ensuremath{\tau}}
\newcommand{\transtimefwd}{\ensuremath{\transtime^-}}
\newcommand{\transtimebwd}{\ensuremath{\transtime^+}}

\newcommand{\measmatrixinv}{\mat{N}}

\newcommand{\btau}{\bar{\tau}}

\newcommand{\odesystemmatrix}{\ensuremath{\mat F}}

\newcommand{\odeinputvector}{\ensuremath{\bm g}}
\newcommand{\odeoutputvector}{\ensuremath{\bm{c}^\intercal_0}}

\newcommand{\pdeobservabilitymatrix}{\ensuremath{\mathcal{O}_H}}
\newcommand{\pdeobservabilityfeedthroughmatrix}{\mat{D}_H}
\newcommand{\pdetimeshiftfwd}[2]{\ensuremath{\gamma^-(#2;#1)}}
\newcommand{\pdetimeshiftbwd}[2]{\ensuremath{\gamma^+(#2;#1)}}

\newcommand{\examplepdestate}{\ensuremath{x}}
\newcommand{\exampleodestate}{\ensuremath{\xi}}
\newcommand{\examplemass}{\ensuremath{m}}
\newcommand{\examplestiffness}{\ensuremath{k}}

\newcommand{\Lp}{L^2}
\newcommand{\LpLoc}{L^2_{\text{loc}}}
\newcommand{\SobLoc}[1]{H^{#1}_{\text{loc}}}
\newcommand{\BV}{\mathrm{BV}}

\newcommand{\kpp}[1][z_0]{k_{#1}^{++}}
\newcommand{\kmm}[1][z_0]{k_{#1}^{--}}
\newcommand{\kpm}[1][z_0]{k_{#1}^{+-}}
\newcommand{\kmp}[1][z_0]{k_{#1}^{-+}}
\newcommand{\kppm}[1][z_0]{k_{#1}^{+\pm}}
\newcommand{\kmpm}[1][z_0]{k_{#1}^{-\pm}}

\newcommand{\mat}[1]{\mathbf{#1}}
\newcommand{\vect}[1]{\bm{#1}}

\newcommand{\domk}{\Gamma_{\text{inf}}(z_0)}

\definecolor{mplblue}{RGB}{31, 119, 180}
\definecolor{mplorange}{RGB}{255, 127, 14}
\definecolor{mplgreen}{RGB}{44, 160, 44}
\definecolor{mplred}{RGB}{214, 39, 40}
\definecolor{mplpurple}{RGB}{148, 103, 189}
\definecolor{mplbrown}{RGB}{140, 86, 75}
\definecolor{mplpink}{RGB}{227, 119, 194}
\definecolor{mplgray}{RGB}{127, 127, 127}
\definecolor{mplolive}{RGB}{188, 189, 34}
\definecolor{mplcyan}{RGB}{23, 190, 207}
\pgfdeclarelayer{bg}
\pgfsetlayers{bg,main}

\tikzstyle{block} = [draw, fill=white, rectangle, minimum height=2em, minimum width=2cm]

\tikzstyle{gain} = []
\tikzstyle{rblock} = [draw, fill=white, rectangle, minimum height=3em, minimum width=5.5cm, text width=5.5cm]
\tikzstyle{nlblock} = [draw, postaction={draw,line width=0.25mm,white}, line width=0.5mm, black, fill=white, rectangle, minimum height=3em, minimum width=5em]
\tikzstyle{sum} = [draw, circle]
\tikzstyle{branch} = [circle,inner sep=0pt,minimum size=1mm,fill=black,draw=black, anchor=mid]
\tikzstyle{nvbranch} = [circle,inner sep=0pt,minimum size=1mm,fill=white,draw=white, fill opacity=0, draw opacity=0]
\tikzstyle{vecBranch} = [circle,inner sep=0pt,minimum size=2mm,fill=black,draw=black]
\tikzstyle{input} = [coordinate]
\tikzstyle{output} = [coordinate]
\tikzstyle{coord} = [coordinate]

\tikzset{%
	gain/.style 2 args={
		draw,
		fill=white,
		isosceles triangle,
		isosceles triangle apex angle=60,
		minimum height = 2.5em,
		outer sep=0,
		text width=2em,
		label={[shift={(-0.125ex,#2)}]center:#1}
	},%
	transport/.style={%
		draw,
		label={},
		path picture={
			\pgfpointdiff{\pgfpointanchor{path picture bounding box}{north east}}%
			{\pgfpointanchor{path picture bounding box}{south west}}
			\pgfgetlastxy\x\y
			\tikzset{x=\x*.4, y=\y*.4}
			\draw[line width=0.25mm] (-0.9,0.3) ellipse (0.15 and 0.4);
			\draw[-, line width=0.1mm] (-0.8825, 0.575) arc [start angle=80, end angle=-240, x radius=0.105, y radius=0.28];
			\draw[->, line width=0.05mm] (-0.91, 0.55) arc [start angle=-230, end angle=-250, x radius=0.105, y radius=0.28];
			\draw[line width=0.25mm] (0.9,0.3) ellipse (0.15 and 0.4);
			\draw[-, line width=0.1mm] (0.9175, 0.575) arc [start angle=80, end angle=-240, x radius=0.105, y radius=0.28];
			\draw[->, line width=0.05mm] (0.89, 0.55) arc [start angle=-230, end angle=-250, x radius=0.105, y radius=0.28];
			\draw[line width=0.25mm, name path=A] (-0.9,-0.1) -- (0.9,-0.1);
			\draw[line width=0.25mm] (-0.9,0.7) -- (0.9,0.7);
			\draw[line width=0.15mm, name path=B] (-0.9,-0.1) to[out=100, in=350] (-0.6,-0.9) to [out=170, in=30] (0, -0.4) to [out=210, in=350] (0.7, -0.6) to [out=170,
			in=60] (0.9, -0.1);
			\node[align=center] at (1.25,0.3) () {#1};
			\begin{pgfonlayer}{bg}
				\draw [left color=black, right color=white, middle color=black!20, intersection segments={of=A and B}];
			\end{pgfonlayer}
		}
	},
}

\newacronym[plural={PDEs}, longplural={partial differential equations}]{pde}{PDE}{partial differential equation}
\newacronym[plural={ODEs}, longplural={ordinary differential equations}]{ode}{ODE}{ordinary differential equation}
\newacronym[plural={BCs}, longplural={boundary conditions}]{bc}{BC}{boundary condition}
\newacronym[plural={FDEs}, longplural={functional differential equations}]{fde}{FDE}{functional differential equation}
\newacronym{hcf}{HCF}{hyperbolic controller form}
\newacronym{ocf}{OCF}{observer canonical form}
\newacronym{hccf}{HCCF}{hyperbolic controller canonical form}
\newacronym{ccf}{CCF}{controler canonical form}
\newacronym{oycf}{OyCF}{observability canonical form}
\newacronym{honf}{HOCF}{hyperbolic observer canonical form}
\newacronym{hoynf}{HOyNF}{hyperbolic observability normal form}
\newacronym{bf}{BF}{Brunovsky form}
\newacronym{siso}{SISO}{single-input single-output}
\newacronym{dde}{DDE}{delay differential equation}
\newacronym{doi}{DOI}{domain of influence}
\newacronym{dod}{DOD}{domain of determinacy}

\begin{document}

\maketitle
\thispagestyle{empty}
\pagestyle{empty}

\begin{abstract}
A hyperbolic observer canonical form (HOCF) for linear hyperbolic PDEs with boundary dynamics is presented. The transformation to the HOCF is based on a general procedure that uses so-called observability coordinates as an intermediate step. These coordinates are defined from an input--output relation given by a neutral functional differential equation (FDE), which, in the autonomous case, reduces to an autonomous FDE for the output. The HOCF coordinates are directly linked to this FDE, while the state transformation between the original coordinates and the observability coordinates is obtained by restricting the observability map to the interval corresponding to the maximal time shift appearing in the FDE. The proposed approach is illustrated on a string--mass--spring example.
\end{abstract}


\section{Introduction}\label{sec:introduction}
Canonical coordinate representations play a central role in the analysis and
observer design of finite-dimensional systems. In particular, controller
and observer canonical forms express the system dynamics directly in terms
of input or output derivatives
\cite{brunovskyControllableSystems1970,zeitzCanonicalFormsNonlinear1989}.
The transfer of these ideas to distributed-parameter systems has been
investigated in various directions. Early work on canonical representations
for hyperbolic distributed-parameter systems can be found in
\cite{russellCanonicalFormsSpectral1978}, where hyperbolic systems were
related to delay-type descriptions.

Hyperbolic distributed-parameter systems arise naturally in applications
such as transport processes, traffic flow, and wave propagation. Their
analysis and control have therefore received considerable attention in the
literature. A large portion of these contributions is devoted to backstepping-based control of distributed-parameter systems; see, for example, the books
\cite{bastinStabilityBoundaryStabilization2016,krsticBoundaryControlPDEs2008} and the review paper \cite{VazquezAuriolBribiescaArgomedoKrstic2025automatica}.
Moreover, structural representations based on flatness and
canonical coordinates have been investigated for hyperbolic systems. In
particular, flatness-based parametrizations and controller canonical forms
have been developed in
\cite{woittennekControllerCanonicalForms2012a, woittennekFlatnessBasedFeedback2013a, woittennekFlatnessbasedAnalysisControl2022}. 
An introduction to such normal-form approaches for control of distributed-parameter systems is provided in \cite{gehringControlDistributedparameterSystems2023}, which also clarifies the connection to the backstepping-based approaches.

Observer design for hyperbolic systems has mainly been addressed using
backstepping techniques, where integral transformations are used to map the
original system to a stable target system
\cite{vazquezBacksteppingBoundaryStabilization2011b, dimeglioBacksteppingBoundaryObserver2013,Aamo2013tac,HasanAamoKrstic2016}. Observer constructions for nonlinear coupled systems have also been considered, for instance in \cite{irscheidObserverDesign2x22021}, where a \gls{pde} is coupled at the boundary to an \gls{ode}, forming a PDE-ODE system.

An alternative viewpoint is provided by observer-oriented canonical
representations. In particular, based on \cite{russellCanonicalFormsSpectral1978}, the hyperbolic
observer canonical form was introduced in \cite{woittennekHyperbolicObserverCanonical2013} by exploiting the relation between hyperbolic
distributed-parameter systems and \glspl{fde} (see also \cite{RiesmeierWoittennek2023ifac,gehringControlDistributedparameterSystems2023}).

The present work applies the approach proposed in \cite{woittennekHyperbolicObserverCanonical2013}
to general \gls{siso} linear hyperbolic PDE--ODE systems. More precisely, two-scalar, heterodirectional transport PDEs are considered, which are bidirectionally coupled  to an ordinary differential equation at one boundary, with the measured output residing at the opposite boundary.
The considered system class is formulated in a standard setting for heterodirectional hyperbolic PDE–ODE systems that is well established in the backstepping literature. In particular, spatially varying in-domain coupling is included in the commonly used off-diagonal form.
Using the method of characteristics, the distributed state is parameterized by the restriction
of the output trajectory to a compact interval, giving rise to the definition of the so-called observability coordinates. The output itself can be shown to satisfy a neutral \gls{fde} which serves as a basis for both the definition of the observer canonical form associated with the given system as well for the computation of the transformation between the observability coordinates and the observer coordinates.

The remainder of the paper is organized as follows:
In Section~\ref{sec:honf} the \gls{honf} and the connection with the observability coordinates are revisited in a general fashion. Afterwards,
in Section~\ref{sec:hyperbolic-systems} the considered class of
hyperbolic PDE--ODE systems is introduced and the general procedure is further detailed for this  system class.
Finally, Section~\ref{sec:example} illustrates the proposed approach using a
string--mass--spring example and 
Section~\ref{sec:conclusion} concludes the paper.

\section{Preliminaries \& Notation}\label{sec:preliminaries}
For $n \in \naturals$, $\reals^n$ and $\reals^{n\times n}$ denote the $n$-dimensional Euclidean space and the set of real-valued $n\times n$ matrices, respectively. 
Given an interval $\Omega \subset \reals$, $\Lp(\Omega,\reals^n)$ denotes the Lebesgue space of square-integrable functions taking values in $\reals^n$, while $H^k(\Omega,\reals^n)$ denotes the corresponding Sobolev space of $k$ times weakly differentiable functions. The corresponding spaces of locally integrable functions on $\reals^+$ are denoted by $\LpLoc(\reals^+,\reals^n)$ and $\SobLoc{k}(\reals^+,\reals^n)$. Moreover, $\BV(\Omega)$ and $V(f,\Omega)$ denote the real-valued functions of bounded variation on the interval $\Omega$ and the total variation of $f\in\BV(\Omega)$ on $\Omega$, respectively. Eventually, for any topological space $\mathcal{M}$, $C_0(\Omega, \mathcal{M})$ denotes the space of continuous functions on $\Omega$ with values in $\mathcal{M}$. 
Partial derivatives of a function $f$ of several variables $z_i$, $i=1,2,\dots$ w.r.t.\ $z_i$
are denoted by $\pderi[z_i]f$. Alternatively, the partial derivative w.r.t. time $t$ is also denoted
using the notation $\partial_t f=\dot f$,  $\partial_t^2 f=\ddot f$,  $\partial_t^i f= f^{(i)}$, $i\in\naturals$. For a function $f$ of two (or more) variables $z_1$ and $z_2$, the notation $f(z_1,\bullet)$ is used in order to indicate that $f$ is considered a function of the second argument $z_2$ only, the first argument being fixed, i.e., to denote the function $z_2\mapsto f(z_1,z_2)$.
For a function $f:\reals^+ \to \reals$, $\vect f^{[n]}(t)\in\reals^{n+1}$ denotes the vector of time derivatives up to order $n$:
$$
\vect f^{[n]}(t) := \bigl(f(t), \dot f(t), \dots, f^{(n)}(t)\bigr)^\intercal.
$$
Furthermore, the shift operator $\timeshiftoperator{\theta}$, $\theta \in \reals$, is defined by $(\rho_\theta f)(t) := f(t+\theta)$.
Finally, the space of bounded linear operators between Banach spaces $\mathcal{M}$ and $\mathcal{N}$ is denoted by $\Lin(\mathcal{M},\mathcal{N})$.

\section{General aspects} \label{sec:honf}
\subsection{Definition of the \glsentryfull{honf}}
\begin{figure}[ht]
    \centering
    \adjustbox{width=\linewidth}{
        \begin{tikzpicture}[auto, >=latex]
	\def\dstyle{dashed} 
	\def\length{1.0cm} 
	\def\intwidth{1ex} 
	\def\vdistance{1.0cm} 
	\def\hdistance{1.0cm} 
	\def\ddistance{1.0cm} 
	
    \node [sum] (msum1) {};
    \node [block, right=\hdistance of msum1, minimum width=\intwidth] (mint1) {\huge$\int$};
    \node [sum, right=\hdistance of mint1] (msum2) {};
    \node [block, right=\hdistance of msum2, minimum width=\length, white, fill=white] (dots) {};
    \node [sum, right=\hdistance of dots] (msum4) {}; 
    \node [block, right=\hdistance of msum4, minimum width=\intwidth] (mint3) {\huge$\int$};
    \node [transport=\footnotesize$\eta_{n+1}$, right=\hdistance of mint3, minimum width=3cm, minimum height=1.05cm] (trans) {};
    \node [sum, opacity=0] (phantom) at (trans){};
    \node [coord, right=\hdistance of trans] (mbranch1) {};
    \node [coord, right=\hdistance of mbranch1] (output) {};
    
    \node [gain={$a_{1}$}{0.5ex}, shape border rotate=90, below=\vdistance of msum2] (aGain2) {};
    \node [gain={$a_{0}$}{0.5ex}, shape border rotate=90, below=\vdistance of msum1] (aGain1) {};
    \node [branch, below=\vdistance of aGain2] (bbranch2) {};
    \node [gain={$a_{n\shortminus1}$}{0.5ex}, shape border rotate=90, below=\vdistance of msum4] (aGainN) {};
    \node [branch, below=\vdistance of aGainN] (bbranchN) {};
    \node [gain={$\shortminus a_{n}$}{0.525ex}, shape border rotate=90, below=\vdistance of phantom] (aGainNp1) {};
    \node [branch, below=\vdistance of aGainNp1] (bbranchNp1) {};
    
	\path (msum2 |- msum2) ++(\ddistance,0) coordinate (mstart);
	\path (mstart |- bbranch2) coordinate (bstart);
	\path (mstart) ++(\length,0) coordinate (mend);
    \path (bstart) ++(\length,0) coordinate (bend);

    \draw[->] (aGain1) -- node[at end, right]{$\shortminus$}(msum1);
    \draw[->] (msum1) -- node[midway, above]{$\dot \eta_1$} (mint1);
    \draw[->] (mint1) -- node [above] {$\eta_1$} (msum2);
    \draw[->] (msum4) -- node[midway, above]{$\dot \eta_{n}$}(mint3);
    \draw[->] (mint3) -- node [above] {$\eta_{n}$} (trans);
    \draw[] (trans) --  node [midway, above] {$y$}(mbranch1);
    \draw[->] (mbranch1) |- (bbranchNp1) -- (aGainNp1);
    \draw[->] (aGainNp1) -- (trans);
    \draw[->] (bbranchNp1) -- (bbranchN) -- (aGainN);
    \draw[->] (aGainN) -- node[at end, right]{$\shortminus$}(msum4);
    \draw[->] (bbranch2) -- (aGain2);
    \draw[->] (aGain2) -- node[at end, right]{$\shortminus$}(msum2);
    \draw[->] (bbranch2) -| (aGain1);
        
    \draw[] (msum2) -- node[midway, above]{$\dot \eta_2$} (mstart);
    \draw[\dstyle] (mstart) -- (mend);
    \draw[->] (mend) -- node[midway, above]{$\eta_{n-1}$}(msum4);
    
    \draw[] (bbranchN) -- (bend);
    \draw[\dstyle] (bend) -- (bstart);
    \draw[-] (bstart) -- (bbranch2);
\end{tikzpicture}
    }
    \caption{Schematic of the \gls{honf}.}
    \label{fig:infdim:honf}
\end{figure}
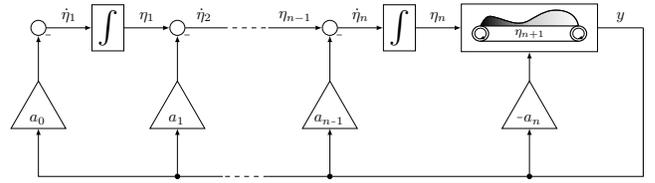

The (linear \gls{siso}) \gls{honf} can be defined as the dual of the \gls{hccf} \cite{gehringControlDistributedparameterSystems2023,RiesmeierWoittennek2023ifac,woittennekHyperbolicObserverCanonical2013,russellCanonicalFormsSpectral1978}. 
Like the finite-dimensional \gls{ocf} it consists of an integrator chain
\begin{subequations}\label{eq:infdim:hocf}
  \begin{align}
    \begin{split}
      \dot \eta_1(t)&=-a_0 y(t)\\
      \dot \eta_i(t)&=\eta_{i-1}(t)-a_{i-1} y(t),\quad i=2,\dots, n
    \end{split}\label{eq:infdim:hocf:ode}
\end{align}
with injection of the system output $y(t)$\footnote{For the sake of brevity in the general part of this contribution only autonomous systems are considered which is not a restriction.}, describing the evolution of the finite-dimensional part of the state which is attached to the input
\begin{align}
                 \eta_{n+1}(0,t)&=\eta_n(t)\label{eq:infdim:hocf:bc}
\end{align}
of the transport \gls{pde}
\begin{align}
    \partial_t{\eta}_{n+1}(\tau,t)&=-\partial_\tau\eta_{n+1}(\tau,t)-a_ny(t),\quad \tau\in[0,\hat\tau]\label{eq:infdim:hocf:pde}
  \end{align}
   with state $\eta_{n+1}(\bullet,t)\in\Lp([0,\hat\tau])$.
  Finally, the measured output variable corresponds to the outflow
\begin{equation}
  y(t)=\eta_{n+1}(\hat\tau,t)\label{eq:infdim:hocf:output}
\end{equation}
of the transport system.
\end{subequations}
A schematic representation of \eqref{eq:infdim:hocf} is shown in Fig.~\ref{fig:infdim:honf}. Above $a_i\in\reals$, $i=0,\dots,n-1$ while the  possibly unbounded distributed output injection operator $a_n$  is defined by
\begin{equation*}
  a_{n}(\tau)=\frac{\dd\alpha}{\dd\tau}(\tau)
\end{equation*}
with $\alpha \in\BV([0,\hat\tau])$  satisfying (cf.\ \cite{russellCanonicalFormsSpectral1978})
\begin{equation*}
    \lim_{\epsilon\rightarrow 0} V(\alpha,[\hat\tau-\epsilon,\hat\tau])=0.
\end{equation*}

\subsection{Input-output-relation and observability coordinates}\label{sec:general:io}
\begin{figure}
    \centering
    \adjustbox{width=\linewidth}{
        \begin{tikzpicture}[
    node distance=3cm,
    >=Latex,
    block/.style={
        rectangle,
        draw,
        rounded corners,
        minimum width=2.5cm,
        minimum height=1.2cm,
        align=center
    },
    trans/.style={->, thick}
]

\node[block] (orig) {%
    original\\system
};

\node[block, right=of orig] (obs) {%
    observability \\coordinates
};

\node[block, right=of obs] (onf) {%
     \gls{honf} 
};

\draw[trans] ([yshift=5pt]orig.east) -- 
node[above] {$\observabilityMap$} 
([yshift=5pt]obs.west);

\draw[trans] ([yshift=5pt]obs.east) -- 
node[above] {$\honftransformation$} 
([yshift=5pt]onf.west);

\draw[trans, <-] (orig.south) -- ++(0,-1) -| (onf.south);
\node[fill=white, inner sep=1pt] at ($(orig.south)!0.5!(onf.south)+(0,-1)$)
{$\invobservabilityMap\circ\invhonftransformation  $};

\draw[trans, <-] (onf.north) -- ++(0,1) -| (orig.north);
\node[fill=white, inner sep=1pt] at ($(orig.north)!0.5!(onf.north)+(0,1)$)
{$\honftransformation \circ \observabilityMap$};

\draw[trans] ([yshift=-5pt]obs.west) -- 
node[below] {$\invobservabilityMap$} 
([yshift=-5pt]orig.east);

\draw[trans] ([yshift=-5pt]onf.west) -- 
node[below] {$\invhonftransformation$}
([yshift=-5pt]obs.east);

\end{tikzpicture}
    }
    \caption{Transformations between the original system and the \gls{honf} via observability coordinates.}
    \label{fig:transformations:observer-forms}
\end{figure}
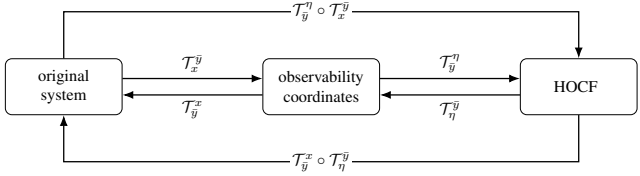

Integrating \eqref{eq:infdim:hocf:pde} along the characteristics and employing the output equation \eqref{eq:infdim:hocf:output} yields the Volterra integral equation
\begin{subequations}\label{eq:infdim:hocf:transform_to_bary}
  \begin{equation}\label{eq:infdim:hocf:volterra}
    \eta_{n+1}(\tau,t+\tau)=y(t+\hat\tau)+\int_{\tau}^{\hat\tau} y(t+\btau)d\alpha(\btau)
  \end{equation}
  relating the distributed part of the state and the restriction of the output trajectory to the interval $[t,t+\hat\tau]$. Note that for any $\eta_{n+1}(\bullet,t)\in \Lp([0,\hat\tau])$, 
  \eqref{eq:infdim:hocf:volterra} admits a unique solution $y([t,t+\hat\tau])\in \Lp([0,\hat\tau])$.

  Evaluating \eqref{eq:infdim:hocf:volterra} at $\tau=0$, plugging the result into \eqref{eq:infdim:hocf:bc},
  and evaluating \eqref{eq:infdim:hocf:ode} yields  
  \begin{equation}\label{eq:infdim:hocf:trafo_findim}
    \eta_{n-i}(t)=y^{(i)}(t+\hat\tau)+(a_{n}^* y^{(i)})(t)+\sum_{j=1}^{i}a_{n-j}y^{(i-j)}(t)
  \end{equation}
\end{subequations} for $i=0,\dots,n-1$, as well as the \gls{fde} 
\begin{equation}\label{eq:infdim:io}
       \sum_{i=0}^{n-1}a_iy^{(i)}(t)+y^{(n)}(t+\hat\tau)+a_{n}^* y^{(n)}(t)=0.
\end{equation}
Therein, the distributed prediction operator $a_{n}^*$ is given by
\begin{align*}
   (a_{n}^*f)(t)&=\int_{0}^{\hat\tau}f(t+\btau)\dd\alpha(\btau).
\end{align*}
Introducing the so-called observability coordinates $\bar y(t):=\bar y(\bullet,t)\in\mathcal{Y}= H^{n}([0,\hat\tau])$
via
\begin{equation*}
  \bar y(\tau,t)=y(t+\tau),\quad \tau\in[0,\hat\tau],
\end{equation*}
equations \eqref{eq:infdim:hocf:transform_to_bary} 
define the transformation
\begin{equation*}
    \bar y(\bullet,t) \mapsto (\eta_1(t),\dots, \eta_n(t), \eta_{n+1}(\bullet,t)),
\end{equation*}
and thus the densely defined map
\begin{equation*}
\honftransformation:\Lp([0,\hat{\tau}])\supset H^{n}([0,\hat\tau])\to \reals^n\times\Lp([0,\hat\tau]).
\end{equation*}
In the sequel, these intermediate coordinates constitute a means for the transformation of a given system to the observer canonical form (see e.g. \cite{gehringControlDistributedparameterSystems2023,woittennekHyperbolicObserverCanonical2013}). For a given (hyperbolic) system on a state space $\statespace$ with associated semigroup $S(t)$ and admissible output-operator $\outop$, the observability map $\outop\semigroup\in \Lin(\statespace,\LpLoc(\reals^+))$ is considered on an appropriate subinterval $[0,\hat\tau]\subset\reals^+$ which is chosen such (if possible) that the observability map turns into a
bijection $\mathcal{T}_{x}^{\bar y}$, at least when considered a map from a dense subspace of the original state space to the dense subspace  $\mathcal{Y}\subset\Lp([0,\hat\tau])$. Afterwards the transform $\mathcal{T}_{x}^{\bar y}$ between the original coordinates and the observability coordinates on the one hand and the transform $\mathcal{T}_{\bar y}^{\eta}$ between the observability coordinates and the observer coordinates on the other hand are composed to obtain an overall transform from the original system description to the \gls{honf} (cf.\ Fig.~\ref{fig:transformations:observer-forms}).

\section{Hyperbolic observer form for PDE-ODE systems}\label{sec:hyperbolic}
This section is devoted to the derivation of the \gls{honf} for a particular class of systems, namely linear \gls{siso} systems consisting of two coupled transport equations attached to a finite-dimensional
boundary system\footnote{This class of systems has been well analyzed from a backstepping point of view. However, a direct transformation of the equations to the \gls{honf} is not known to the authors.}. The transformation is achieved by directly mapping the original coordinates to the observability coordinates introduced within the previous section. The subsequent transformation into the 
\gls{honf} coordinates is straightforward.
\subsection{System class}\label{sec:hyperbolic-systems}
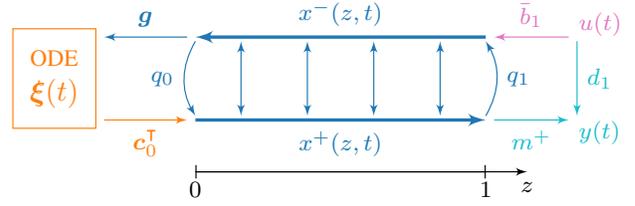
\begin{figure}[!ht]
    \centering
    \adjustbox{width=\linewidth}{
        \begin{tikzpicture}[auto, >=latex']
    \draw[->] (0,-0.125) -- (4, -0.125) node [below] {\footnotesize$z$};
    \draw[|-|] (0,-0.125) node [below] {\footnotesize$0$} -- (3.5, -0.125) node [below] {\footnotesize$1$};

    \draw[->, line width=0.4mm, mplblue] (0,0.5) -- node [below] {\footnotesize$\transstatebwd(z,t)$} (3.5, 0.5);
    \draw[<-, line width=0.4mm, mplblue] (0,1.5) -- node [above] {\footnotesize$\transstatefwd(z,t)$} (3.5, 1.5);

    \draw[->, mplblue] (3.5, 0.55) to [bend right] node [midway, right] {\footnotesize$q_1$} (3.5, 1.45);
    \draw[<-, mplblue] (0, 0.55) to [bend left] node [midway, left] {\footnotesize$q_0$} (0, 1.45);

    \foreach \j in {0.55, 1.35, ..., 3.05 } {
        \draw[<->, mplblue] (\j,0.55) -- (\j, 1.45);
    }

    \draw[->, mplpink] (4.5, 1.5)
      -- node[midway, above] {\footnotesize$\bar b_1$} (3.6, 1.5);

    \draw[->, mplcyan] (4.6, 1.45)
      -- node[midway, right] {\footnotesize$d_1$} (4.6, 0.55);

    \draw[->, mplcyan] (3.6, 0.5)
      -- node[midway, below] {\footnotesize$m^+$} (4.5, 0.5);

    \node[anchor=west, mplpink] at (4.5,1.65) {\footnotesize$u(t)$};
    \node[anchor=west, mplcyan] at (4.5,0.35) {\footnotesize$y(t)$};
    
    \draw[->, mplblue] (-0.1, 1.5) -- node[midway, above] {\footnotesize$\odeinputvector$}(-1.1, 1.5);
    \draw[<-, mplorange] (-0.1, 0.5) -- node[midway, below] {\footnotesize$\odeoutputvector$}(-1.1, 0.5);

    \draw[draw, mplorange] (-2.2,0.4) rectangle (-1.2, 1.6)
      node[pos=.5, anchor=center, align=center] {\footnotesize \acrshort{ode} \\ $\odestate(t)$};
\end{tikzpicture}
    }
    \caption{Visualization of the linear \gls{pde}-\gls{ode} system class.}
    \label{fig:original-system:linear:indomain}
  \end{figure}

The considered system is a vector-valued first-order hyperbolic \gls{pde}
with spatially dependent coefficients
\begin{subequations} \label{eq:original-system}
\begin{equation} \label{eq:original-system:pde}
    \pderi[z] \pdestate(z,t) + \Sigma(z) \pderi[t]\pdestate(z,t) = A(z) \pdestate(z,t)
\end{equation}
defined for $t \in \reals$ and $z \in \Omega = [0,1]$ with 
\begin{align*}
    \Sigma(z) &= \begin{pmatrix}-\sigma^-(z) & 0 \\ 0 & \sigma^+(z)   \end{pmatrix}, & A(z) &= \begin{pmatrix}0 & \mu^-(z) \\ \mu^+(z) & 0 \end{pmatrix},
\end{align*}
$\sigma^\mp\in C_0(\Omega,\reals^+)$, $\mu^\mp\in C_0(\Omega,\reals)$, 
and the distributed state $\pdestate(\bullet,t) = \left(\transstatefwd(\bullet,t),\, \transstatebwd(\bullet,t)\right)^{\intercal} \in \Lp(\Omega,\reals^2)$.
The structural assumptions towards $\Sigma(z)$ and $A(z)$ are imposed without loss of generality (cf. \cite{woittennekFlatnessbasedAnalysisControl2022}, Remark 1). 
At the boundary $z=0$, the \gls{pde} \eqref{eq:original-system:pde} is coupled with the linear observable \gls{siso} system
\begin{align}
    \dot{\odestate}(t) &=  \odesystemmatrix \odestate(t) + \odeinputvector \transstatefwd(0,t)\label{eq:original-system:ode}\\
    \transstatebwd(0,t) &= \odeoutputvector \odestate(t) +  q_0 \transstatefwd(0,t),\label{eq:original-system:ode:output} 
\end{align}
where $\odesystemmatrix \in \reals^{n\times n},\,\odestate(t),\odeinputvector, \odeoutputvector \in \reals^n,\, q_0 \in \reals$.
In order to guarantee exact observability of the overall system $q_0\ne 0$ is assumed.
Moreover, the pair $(\odesystemmatrix,\odeoutputvector)$ is assumed to be in observability canonical form. In the sequel, the last row
of the companion matrix $\odesystemmatrix$ is denoted by $-\vect{f}^\intercal$.

The  model is completed by the \gls{bc}
\begin{align}
    \transstatefwd(1,t) &= q_1 \transstatebwd(1,t) +\bar b_1 u(t), \label{eq:original-system:bc1}
\end{align}
with $q_1,\bar{b}_1 \in \reals$, and output equation
\begin{equation}\label{eq:original-system:output}
    y(t) = m^+ \transstatebwd(1,t) + d_1 u(t),
\end{equation}
\end{subequations}
with feedthrough $d_1\in\reals$.
The overall structure of the model \eqref{eq:original-system} is visualized in Fig.~\ref{fig:original-system:linear:indomain}.
\subsection{State-space description}\label{sec:hyperbolic:state_space}
In an abstract state-space setting the model \eqref{eq:original-system} can be written as
\begin{subequations}\label{eq:original-system:abstract}
  \begin{align}
    \dot{\systemstate}(t) &= \sysop \systemstate(t) + \inpop u(t)\label{eq:system:original-system:abstract:de}\\
    y(t) &= \outop \systemstate(t)\label{eq:system:original-system:abstract:out}.
  \end{align}
\end{subequations}
Therein, the unbounded system operator $\sysop:\statespace\supset \domain(\sysop)\to\statespace$ on the state-space
\begin{equation*}
    \statespace=\pdestatespace\times\odestatespace,\quad  \pdestatespace=\Lp(\Omega, \reals^2),
\end{equation*}
is defined by
\begin{equation}
  \sysop(\bar{\pdestate},\bar{\odestate})=(-\Lambda\partial_z \bar\pdestate+\bar A\bar{\pdestate}, \odesystemmatrix\bar{\odestate} + \odeinputvector\btransstatefwd(0))
\end{equation}
with $\Lambda(z)=(\Sigma(z))^{-1}$,
  $\bar A(z)=\Lambda(z)A(z)$ and domain
\begin{multline*}
  \domain(\sysop)=\Bigl\{(\bar\pdestate,\bar\odestate)\in H_1(\Omega,\reals^2)\times\reals^n|\\ \transstatebwd(0)= \odeoutputvector \odestate +  q_0 \transstatefwd(0),    \transstatefwd(1)= q_1 \transstatebwd(1) 
\Bigr\}.
\end{multline*}
Moreover, the input operator $\inpop\in \statespace_{-1}$ is given by
\[
  \inpop=\left(0,\bm{0}_n,\bar b_1\delta_1\right).
\]
where $\delta_1$ denotes the Dirac distribution centered at $1$ and $\statespace_{-1}$ is the usual extrapolation space (cf.\ \cite{TucsnakWeiss2009}). The following considerations are valid for any (unbounded) output operator $\outop:\statespace\supset\domain(\outop)\to\reals$ with $\domain(\outop)\supset\domain(\sysop)$. A more specific setting is considered later.

Using the method of characteristics in connection with standard fixed-point theory, the following simple result can be shown:
\begin{proposition}\label{prop:well_posedness}
  For any $\bar{\systemstate}\in\statespace$ and any $u\in\LpLoc(\reals^+)$
  \eqref{eq:original-system} (resp.\ \eqref{eq:original-system:abstract}) with initial conditions $\systemstate(0)=\bar{\systemstate}$ possesses a unique mild solution $\systemstate\in C_0(\reals^+,\statespace)$.
Moreover, $\pdestate\in C_0(\Omega,\LpLoc(\reals^+,\reals^2))$. 
\end{proposition}
In the abstract setting, the solution can be written in the form 
\[
    \systemstate(t)=\semigroup(t)\bar{\systemstate}+\reachabilitymap(t)u
\]
where $\semigroup(t)$ denotes the (extrapolated) semigroup generated by $\sysop$ while
$\reachabilitymap(t)=\int_0^t \semigroup(t- s) \inpop\dd s $
is the corresponding reachability map.

\subsection{Parametrization by boundary values}\label{sec:param-bound-valu}
Integrating the transport equations in \eqref{eq:original-system:pde} along their respective characteristic projections
\begin{equation}\label{eq:original-system:characteristics}
  \gamma^\mp(z;z_0)=\int_{z_0}^z\sigma^{\mp}(\zeta)\dd \zeta,
\end{equation}
their solution can be shown to be parameterizable by the trace at $z=z_0$:
  \begin{equation}\label{eq:original-system:pde:solution:z0}
    \pdestate(z,\bullet) = (\Delta_{z_0}(z)+\bm{\mathcal{K}}_{z_0}(z))\pdestate(z_0,\bullet)
  \end{equation}
  where
  \[
     (\Delta_{z_0}(z)\pdestate(z_0,\bullet))(t)=
     \begin{pmatrix}
       \transstatefwd(z_0,t+\pdetimeshiftfwd{z_0}{z})\\
       \transstatebwd(z_0,t-\pdetimeshiftbwd{z_0}{z})
     \end{pmatrix}
   \]
   and the matrix-valued convolution operator
   \[
     \bm{\mathcal{K}}_{z_0}(z)=
     \begin{pmatrix}
       \bm{\mathcal{K}}_{z_0}^{-}(z)\\
       \bm{\mathcal{K}}_{z_0}^{+}(z)
     \end{pmatrix}
   \]
   defined by
  \[
     (\bm{\mathcal{K}}_{z_0}(z)\pdestate(z_0,\bullet))(t)=\int_{-\pdetimeshiftfwd{z_0}{z}}^{\pdetimeshiftbwd{z_0}{z}}\mathbf{k}_{z_0}(z,\tau)\pdestate(z_0, t-\tau) \dd\tau.
   \]
Therein, the continuous kernel functions in 
   \[
\mathbf{k}_{z_0}=
\begin{bmatrix}
  k^{--}_{z_0}&  k_{z_0}^{-+}\\
  k^{+-}_{z_0}&  k^{++}_{z_0}
\end{bmatrix} \in C_0(\domk,\reals^{2\times2})
\]
are defined on
\[
  \domk=\{(z,\tau)\in\Omega\times\reals|-\pdetimeshiftfwd{z_0}{z}<\tau<\pdetimeshiftbwd{z_0}{z}\}
\]
as the unique solution of the \gls{pde}
\begin{align*}
\frac{\partial \kppm}{\partial z} (z,\tau )+  \sigma^{+}(z ) \frac{\partial\kppm}{\partial \tau} (z,\tau )&=\mu^+(z)\kmpm(z,\tau)\\
\frac{\partial \kmpm}{\partial z} (z,\tau )- \sigma^{-}(z ) \frac{\partial\kmpm}{\partial \tau} (z,\tau )&=\mu^-(z)\kppm(z,\tau)
\end{align*}
with \gls{bc}
\begin{align*}
  \kpp(z,- \gamma^{-}(z,z_{0} ) )&=\kmm(z,\gamma^{+}(z,z_{0} ) )=0\\
  \kpm(z,- \gamma^{-}(z,z_{0} ))  &= \frac{\mu^{+}(z )}{(\sigma^{+}(z ) +\sigma^{-}(z ))}\\
  \kmp(z,\gamma^{+}(z,z_{0} ) ) &= \frac{\mu^{-}(z )}{(\sigma^{+}(z ) +\sigma^{-}(z )) }.
\end{align*}
Note that $\domk$ corresponds to the so-called \gls{doi} of the point $(z_0,0)$ according to \cite[p.~438]{CourantHilbert1953b}.
\subsection{Input-output relation and observability coordinates}\label{sec:hyperbolic:observability}

Using the \gls{bc} \eqref{eq:original-system:bc1} and the output equation \eqref{eq:original-system:output}, the boundary values at $z=1$ can be expressed by
\begin{align}
      \pdestate(1,t) &=  \measmatrixinv\begin{pmatrix}y(t) \\ u(t) \end{pmatrix}   \label{eq:infinite:parameterization:x1_by_yu}
\end{align}
with
\begin{equation*}
    \measmatrixinv =\begin{pmatrix}\bm n^{-\intercal}\\\bm n^{+\intercal}\end{pmatrix}=\begin{pmatrix}n_1^-&n_2^-\\n_1^+ & n_2^+\end{pmatrix}=\frac{1}{m^+}\begin{pmatrix}q_1& \bar b_1 m^+ -q_1 d_1 \\ 1&-d_1\end{pmatrix}.
\end{equation*}
Evaluating  \eqref{eq:original-system:pde:solution:z0} for $z_0=1$ yields 
\begin{align}\label{eq:infinite:parameterization:x_by_yu}
x^{\mp}(z,\bullet)       &= \bm n^{\mp\intercal} \begin{pmatrix}y(\bullet\mp\gamma^{\mp}(1;z)) \\ u(\bullet\mp\gamma^{\mp}(1;z)) \end{pmatrix}\hspace{-0.25em}+
                          \bm{\mathcal{K}}_1^{\mp}(z)\measmatrixinv\begin{pmatrix}y \\ u \end{pmatrix}
\end{align}
in view of \eqref{eq:infinite:parameterization:x1_by_yu} or
\begin{align}\label{eq:infinite:boundary0:parameterization}
    x^{\mp}(0,\bullet)       &= \bm n^{\mp\intercal} \begin{pmatrix}y(\bullet\mp\tau^{\mp}) \\ u(\bullet\mp\tau^{\mp}) \end{pmatrix}+
                          \bm{\mathcal{K}}_1^{\mp}(0)\measmatrixinv\begin{pmatrix}y \\ u \end{pmatrix}
\end{align}
at the boundary $z=0$ with  $\transtime^\mp = \gamma^\mp(1;0)$.
Concerning the boundary \gls{ode}, successively differentiating \eqref{eq:original-system:ode:output} and using \eqref{eq:original-system:ode} yields
\begin{equation}\label{eq:ode:ode}
    \pdestate^{+[n-1]}(0,t) =
    \odestate(t) + \pdeobservabilityfeedthroughmatrix \pdestate^{-[n-1]}(0,t)
  \end{equation}
as well as
\begin{equation}\label{eq:ode:io}
  \hat{\vect{f}}^\intercal\pdestate^{+[n]}(0,t)=\hat{\vect{g}}^\intercal\pdestate^{-[n]}(0,t)
  \end{equation}
with
 \begin{align*}
   \hat{\vect{f}}^\intercal=\begin{pmatrix}\vect{f}^\intercal&1  \end{pmatrix},\quad
   \hat{\vect{g}}^\intercal=\begin{pmatrix}\vect{f}^\intercal \pdeobservabilityfeedthroughmatrix&0\end{pmatrix}+\begin{pmatrix}
    g_{n}&\cdots&g_1&q_0  
  \end{pmatrix},
  \end{align*}
and
\begin{equation*}
    \pdeobservabilityfeedthroughmatrix = \begin{pmatrix}
        q_0  & 0 & \cdots & 0 \\
        g_1 & q_0 & \cdots & 0 \\
        \vdots  &\ddots & \ddots & \vdots \\
        g_{n-1} & \cdots & g_1 & q_0
    \end{pmatrix}.
  \end{equation*}
  Substituting \eqref{eq:infinite:boundary0:parameterization} into \eqref{eq:ode:ode} and \eqref{eq:ode:io} with $u(t)=0$ yields 
\begin{align}
  \begin{split}
    \odestate &= 
(n_1^+\timeshiftoperator{\transtimebwd}+\bm{\mathcal{K}}_1^{+}(0)\vect{n}_1) \vect{y}^{[n-1]}-\\ & \qquad \pdeobservabilityfeedthroughmatrix  (n_1^-\timeshiftoperator{-\transtimefwd}+\bm{\mathcal{K}}_1^{-}(0)\vect{n}_1)\vect{y}^{[n-1]}  
  \end{split}
\label{eq:infinite:ode:parameterization}
\end{align}
and 
\begin{multline} \label{eq:infinite:dde:unsorted}
  \left(n_1^+\timeshiftoperator{\transtimebwd}+\bm{\mathcal{K}}_1^{+}(0)\vect{n}_1\right)\hat{\vect{f}}^\intercal\vect{y}^{[n]} =\\
    \left( n_1^-\timeshiftoperator{-\transtimefwd}+\bm{\mathcal{K}}_1^{-}(0)\vect{n}_1\right)\hat{\vect{g}}^\intercal\vect{y}^{[n]}.
\end{multline}
Predicting \eqref{eq:infinite:dde:unsorted} by $\transtimefwd$ and applying multiple integrations by parts, \eqref{eq:infinite:dde:unsorted} can be rewritten in the form
\eqref{eq:infdim:io}, directly enabling the introduction of the observer coordinates $\vect{\eta}$ in terms of the restriction $\bar y(\tau,t)=y(t+\tau)$, $\tau\in[0,\hat\tau]$  of the output trajectory to the interval $[t,t+\hat\tau]$ according to Section \ref{sec:honf}. 
\subsection{State transform}\label{sec:hyperbolic:transform}
It remains to compute the transformation between the original coordinates $\systemstate(t)=(\pdestate(\bullet,t),\odestate(t))$ and the observer coordinates, which is only shortly sketched in the sequel. As the transformation between the observer coordinates and the observability coordinates has already been established in Sec.~\ref{sec:honf}, only the transformation between the original coordinates
and the observability coordinates remains to be computed. The direction $\statespace\ni\systemstate\mapsto \bar{y}(\bullet,t)\in\Lp([0,\hat\tau])$ is straightforward according to Prop.~\ref{prop:well_posedness}, i.e.,
\[
  \bar{y}(\tau,t)=\mathcal{C}S(\tau)\systemstate(t),\quad \tau\in[0,\hat\tau].
\]
The inverse mapping, however, is more involved. Given $\bar{y}(\bullet,t)$, in a
first step $\systemstate(t+\tau^-)$ is computed. This can be easily achieved by
evaluating appropriately shifted versions of
\eqref{eq:infinite:ode:parameterization} and \eqref{eq:infinite:parameterization:x_by_yu} with a shift by $\transtimefwd$. In a second step, the system equations
\eqref{eq:original-system:pde}, \eqref{eq:original-system:ode}, \eqref{eq:original-system:ode:output} with boundary conditions (cf.\ \eqref{eq:original-system:output})
\begin{equation}
    \transstatebwd(1,t)=\frac1{m^+}{y}(t),\quad 
\end{equation}
are solved backward in time on the interval $[t,t+\transtimefwd]$. The latter equations are well-posed on $\statespace$ as long as $m^+$ and $q_0$ are nonzero.

Alternatively, equation \eqref{eq:infinite:dde:unsorted} can be solved backward in time on the interval $[t-\transtimefwd,t]$ to obtain the restriction of $y$ to the interval
$[t-\transtimefwd,t+\transtimebwd]$ with initial conditions $y(t+\tau)=\bar y(\tau, t)$, $\tau\in[0,\hat\tau]$. This requires the additional assumption $q_1=0$. Subsequently, \eqref{eq:infinite:ode:parameterization} and \eqref{eq:infinite:parameterization:x_by_yu} can be directly evaluated to obtain $\systemstate(t)$.

Regardless of which of the above alternatives is used, the computations involve derivatives of $\bar{y}$ up to order $n$ including point evaluation of derivatives up to order $n-1$ in \eqref{eq:infinite:ode:parameterization}. Therefore, they are only well-defined for $\bar{y}(\bullet,t)\in H^{n}([0,\hat\tau])$ resulting in $\pdestate(\bullet,t)\in H^n(\Omega,\reals^2)$ and $\eta_n\in H^n([0,\tau])$. However, it can be shown that the composed mapping $\vect{\eta}(t)\mapsto\systemstate(t)$ can be continued to the complete state space. The details of this result are omitted here for lack of space and postponed to a forthcoming publication.

\section{Example} \label{sec:example}
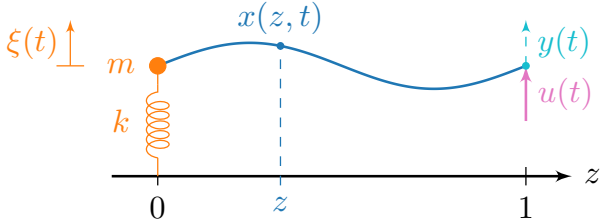
\begin{figure}[ht]
    \centering
    \adjustbox{width=\linewidth}{
        \begin{tikzpicture}[auto, >=latex']

\def\L{4.0}          
\def\zMass{0.0}      
\def\zRight{\L}      
\def\hMass{1.2}      
\def\lSpring{0.8}    
\def\zState{\L/3}    
\def\A{0.25}         
\pgfmathsetmacro{\yState}{\hMass + \A*sin(deg(pi*(\zState)/\L))}

\draw[->,thick] (-0.5,0) -- (\L+0.5,0) node[right] {$z$};

\draw[] (\zMass,-0.1) -- (\zMass,0.0); 
\draw[] (\zRight,-0.1) -- (\zRight,0.1); 

\draw[mplorange,decorate,decoration={coil,aspect=0.55,segment length=3pt,amplitude=3.5pt}]
  (\zMass,\hMass/2 - \lSpring/2) -- (\zMass,\hMass/2 + \lSpring/2);
\draw[mplorange] (\zMass,0)--(\zMass,\hMass/2 - \lSpring/2) ;
\draw[mplorange] (\zMass,\hMass/2 + \lSpring/2)--(\zMass,\hMass) ;
\node[mplorange] at (\zMass-0.4,\hMass/2) {$\examplestiffness$};

\draw[mplorange] (\zMass-1.1,\hMass) -- (\zMass-0.8,\hMass); 
\draw[->,mplorange] (\zMass-0.95, \hMass) --node[midway, left] {$\exampleodestate(t)$}(\zMass-0.95, \hMass+0.5);

\draw[mplblue,line cap=round,thick,domain=0:\L,samples=120,smooth,variable=\zz]
  plot ({\zz},{\hMass + \A*sin(deg(2*pi*\zz/\L))});

\draw[thin,mplblue, dashed] (\zState,-0.1) -- (\zState,\yState);
\filldraw[mplblue] (\zState,\yState) circle(1pt);

\node[below,thin, mplblue] at (\zState,-0.1) {$z$};
\node[above,thin, mplblue] at (\zState,\yState) {$\examplepdestate(z,t)$};

\filldraw[mplorange] (\zMass,\hMass) circle (2.5pt);
\node[mplorange] at (\zMass-0.4,\hMass) {$\examplemass$};

\draw[->, mplpink, thick] (\zRight,\hMass/2) -- (\zRight,\hMass) node[midway,right] {$u(t)$};

\filldraw[mplcyan] (\zRight,\hMass) circle (1pt); 
\draw[->, mplcyan, dashed] (\zRight,\hMass) -- (\zRight,\hMass+0.5) node[midway,right] {$y(t)$};

\node[below] at (\zMass,-0.1) {$0$};
\node[below] at (\zRight,-0.1) {$1$};

\end{tikzpicture}
    }
    \caption{Example of a string with mass-spring boundary.}
    \label{fig:example}
\end{figure}

Consider the transverse displacement $\examplepdestate(z,t)$ of a taut string on the spatial domain $z\in\left[0,1\right]$ and time $t\geq0$, cf. Fig.~\ref{fig:example}.
The in-domain dynamics are governed by the wave equation
\begin{equation}\label{eq:example:secondorder:wave}
    \pderi[z]^2 \examplepdestate(z,t) - \pderi[t]^2 \examplepdestate(z,t) = 0.
\end{equation}
At the left boundary $z=0$, the string is coupled to a mass-spring system with displacement $\exampleodestate(t)$.
The force exerted by the string on the mass equals the boundary traction, which yields the boundary \gls{ode}
\begin{equation}\label{eq:example:secondorder:ode}
    \examplemass \ddot \exampleodestate(t) + \examplestiffness\exampleodestate(t) = - \pderi[z]\examplepdestate(0,t),
\end{equation}
where $\examplemass>0$ and $\examplestiffness>0$ denote the lumped mass and stiffness, respectively. 
Kinematic compatibility at the attachment point imposes that the boundary velocity of the string equals the mass velocity, i.e.
\begin{equation*}
    \pderi[t]\examplepdestate(0,t) = \dot\exampleodestate(t).
\end{equation*}
At the right boundary $z=1$, a control force $u(t)$  is applied to the string yielding the Neumann \gls{bc} 
\begin{equation*}
    \pderi[z]\examplepdestate(1,t) = u(t),
\end{equation*}
while the collocated measurement corresponds to the boundary velocity at $z=1$:
\begin{equation*}
    y(t) = \pderi[t]\examplepdestate(1,t).
  \end{equation*}
Introducing the Riemann coordinates via the transformation
\begin{align*}
  \transstatefwd(z,t)&=\pderi_z x(z,t)+\pderi_t x(z,t)\\
  \transstatebwd(z,t)&=\pderi_z x(z,t)-\pderi_t x(z,t)
\end{align*}
with inverse
\begin{equation*}
  x(z,t)=\xi(t)+\int_0^z\frac{1}{2}\left(\transstatefwd(z,t)+\transstatebwd(z,t)\right)dz,
\end{equation*}
the wave equation \eqref{eq:example:secondorder:wave} can be rewritten as two counter-propagating transport equations
\begin{subequations}\label{eq:example:system}
\begin{align} 
    \pderi_z \transstatefwd(z,t) - \pderi_t \transstatefwd(z,t) &= 0,\label{eq:example:system:pdebwd} \\
    \pderi_z \transstatebwd(z,t) + \pderi_t \transstatebwd(z,t) &= 0.\label{eq:example:system:pdefwd}
\end{align}
Moreover, with $\xi_1(t) = \xi(t)$ and $\xi_2(t) = \dot \xi(t)$, the boundary \gls{ode} \eqref{eq:example:secondorder:ode} appears in the form: 
\begin{align}\label{eq:example:system:ode:xi1}
    \dot \xi_1(t) &= \xi_2(t), \\
    \dot \xi_2(t) &= -\frac{k}{m} \xi_1(t)+\frac{1}{m}\xi_2(t) -\frac1m \transstatefwd(0,t).\label{eq:example:system:ode:xi2}
\end{align}
The system is completed by the \glspl{bc}
\begin{align}
    \transstatebwd(0,t) &= - 2 \xi_2(t) + \transstatefwd(0,t),\label{eq:example:bc0}\\
    \transstatefwd(1,t) &= - \transstatebwd(1,t) + 2 u(t), \label{eq:example:bc1}
\end{align}
and the collocated measurement via 
\begin{equation}
    y(t) = \frac12 \transstatefwd(1,t) - \frac12 \transstatebwd(1,t)= -\transstatebwd(1,t)+u(t). \label{eq:example:meas}
\end{equation}
\end{subequations}
For the sake of completeness and to apply the presented methodology accordingly, the system matrices, vectors and multipliers are given by
\begin{align*}
    \odesystemmatrix &= \begin{pmatrix}0&1\\-\frac{k}{m}& \frac{1}{m}\end{pmatrix}, & \odeinputvector &= \begin{pmatrix}0\\-\frac{1}{m}\end{pmatrix}, & \odeoutputvector &= \begin{pmatrix}0 & -2\end{pmatrix},
\end{align*}
and
\begin{align*}
    q_0 &= 1, & q_1 &= -1, & \bar b_1 &= 2, & m^+ &= -1, & d_1 &= 1.
\end{align*}
\subsection{Input-output relation and observability coordinates}
The measurement at the right boundary \eqref{eq:example:meas} and the \gls{bc}~\eqref{eq:example:bc1} are used to parameterize the boundary values via the measurement and the input as
\begin{align}
    \pdestate(1,t) &= \measmatrixinv \begin{pmatrix}y(t)\\u(t)\end{pmatrix}=\begin{pmatrix}1&1\\-1&1\end{pmatrix} \begin{pmatrix}y(t)\\u(t)\end{pmatrix}\label{eq:example:param:x1}.
\end{align}
As in the general part, $u(t)=0$ is assumed in the sequel.
According to \eqref{eq:original-system:pde:solution:z0} and \eqref{eq:infinite:parameterization:x_by_yu} the solutions of the transport equations \eqref{eq:example:system:pdebwd}--\eqref{eq:example:system:pdefwd} are given by
\begin{subequations}\label{eq:example:param:x_by_y}
  \begin{align}
    \transstatefwd(z,t) &= \transstatefwd(1,t+z-1)=y(t+z-1)\\
    \transstatebwd(z,t) &= \transstatebwd(1,t-z+1)=-y(t-z+1)
  \end{align}
\end{subequations}
Evaluating these expressions at $z=0$ and using \eqref{eq:example:param:x1} yields the boundary values
\begin{align}\label{eq:example:param:x0}
    \pdestate(0,t) &= \begin{pmatrix}\transstatefwd(1,t-1) \\ \transstatebwd(1,t+1)\end{pmatrix} =\begin{pmatrix}y(t-1)\\-y(t+1)\end{pmatrix} ,
\end{align}
which is fully parameterized by the measured output $y(t)$ for vanishing input.
To match the structure of Section~\ref{sec:hyperbolic}, the lumped subsystem \eqref{eq:example:system:ode:xi1}--\eqref{eq:example:system:ode:xi2} is expressed in observability coordinates
\begin{equation}
    \odeobservabilitystate(t) := \pdeobservabilitymatrix \odestate(t),\quad \pdeobservabilitymatrix = \begin{pmatrix}0&-2\\\frac{2k}{m} & -\frac{2}{m}\end{pmatrix}.\label{eq:example:ode:observability}
\end{equation}
Subsequently, substituting \eqref{eq:example:param:x0} into \eqref{eq:infinite:ode:parameterization} yields the \gls{ode} state in terms of the output trajectory
\begin{equation} 
    \odeobservabilitystate(t) = \vect{y}^{[1]}(t+1)-\pdeobservabilityfeedthroughmatrix \vect{y}^{[1]}(t-1)\label{eq:example:param:xi:general}, \, \pdeobservabilityfeedthroughmatrix = \begin{pmatrix}1&0\\\frac{2}{m}&1\end{pmatrix}.
\end{equation}
Inverting \eqref{eq:example:ode:observability} results in the explicit relations
\begin{subequations} \label{eq:example:param:xi}
\begin{align}
    \begin{split}
    \xi_1(t) &= -\frac{m}{2k}\left(\dot y(t-1) + \dot y(t+1)\right) +\\
             &\hphantom{=}+\frac{1}{2k}\left(y(t+1) - y(t-1)\right), 
\end{split}\\
        \xi_2(t) &= \frac12\left( y(t+1)+y(t-1)\right).
   \end{align}
\end{subequations}
To derive the input-output relation and eliminate the internal states, one computes the second time derivative of~\eqref{eq:example:bc0}  and substitutes the expressions \eqref{eq:example:param:xi}. This yields the \gls{fde} 
\begin{align} \label{eq:example:dde}
    \begin{split}
        0 &= \ddot y(t+1) - \frac1m\dot y(t+1) + \frac{k}{m} y(t+1) + \\
          &\hphantom{=} +\ddot y(t-1) + \frac1m \dot y(t-1) + \frac{k}{m}y(t-1).
    \end{split}
\end{align}
In order to introduce the observer coordinates, \eqref{eq:example:dde} is shifted in time
\begin{equation}\label{eq:example:dde:shifted}
    0 = \ddot y(t+2) - \frac1m \dot y(t+2) + \frac{k}{m} y(t+2) + \ddot y(t) + \frac1m \dot y(t) +\frac{k}{m}y(t).
\end{equation}
According to Section \ref{sec:general:io}, the required initial conditions for $y$ on the interval $[t,t+2]$ correspond to the observability coordinates $\bar y$.

In order to introduce the observer coordinates, the shifted derivatives are rewritten using the identity
\begin{equation*}
    y^{(i)}(t+\hat{\tau}) = y^{(i)}(t) + \int_0^{\hat{\tau}} y^{(i+1)}(t+\btau)\dd\btau
\end{equation*}
which follows from the Newton-Leibniz formula. Applying this relation repeatedly transfers \eqref{eq:example:dde:shifted} into the form \eqref{eq:infdim:io}:
\begin{align}\label{eq:example:dde:shifted:rewritten}
    \begin{split}
        0 &= \ddot y(t+2) + \ddot y(t)+ \frac{2k}{m} \dot y(t)+\frac{2k}{m} y(t) + \\
          &\hphantom{=} +\int_0^2 \left(\frac{k}{m}(2-\btau)-\frac1m\right)\ddot y(t + \btau) \dd\btau.
    \end{split}
\end{align}
Since the integral kernel in \eqref{eq:example:dde:shifted:rewritten} is continuous, the corresponding measure is absolutely continuous, while the term $\ddot y(t)$ corresponds to an atom of weight one at $\btau=0$. Hence, the distributed term can be represented as the Stieltjes operator
\begin{equation*}
    (a_2 \ddot y)(t) := \int_0^2 \ddot y(t+\btau) \dd\alpha(\btau),
\end{equation*}
where the density of the measure is given by
\begin{equation*}
   \alpha(0)=1,\quad \alpha^\prime(\tau) =\frac{k}{m}(2-\tau)-\frac1m,\quad \tau \in \left[0,2\right],
\end{equation*}
and consequently,
\begin{equation*}
    (a_2 \ddot y)(t) =\ddot y(t)+ \int_0^2 \left(\frac{k}{m}(2-\btau)-\frac1m\right)\ddot y(t+\btau)\dd\btau.
\end{equation*}
\subsection{\Glsentryfull{honf}}
The observer coordinates and the corresponding \gls{honf} are now introduced according to Section \ref{sec:general:io}.
To this end, \eqref{eq:example:dde:shifted:rewritten} is written in the nested form
\begin{equation}\label{eq:example:honf:fde:ausgeklammert}
    0 = \frac{2k}{m} y(t) + \derifr{}{t}\left(\frac{2k}{m} y(t) + \derifr{}{t}\left(y(t+2)+(a_2y)(t)\right)\right).
\end{equation}
The lumped observer coordinates are now introduced as the
parenthesized expressions in \eqref{eq:example:honf:fde:ausgeklammert}, namely
\begin{subequations} \label{eq:example:honf:trafo}
\begin{align}
    \eta_1(t) &= \frac{2k}{m}y(t) + \dot\eta_2(t)\label{eq:example:honf:trafo:1}\\
    \eta_2(t) &= y(t+2)+(a_2y)(t).\label{eq:example:honf:trafo:2}
\end{align}
The delayed output contribution contained in $\eta_2(t)$ is 
represented by the distributed coordinate (cf.\ \eqref{eq:infdim:hocf:volterra})
\begin{equation}
    \eta_3(\theta,t) = y(t+2-\theta)+\int_\theta^2 y(t+\btau-\theta)\dd\alpha(\btau),
\end{equation}
\end{subequations}
which satisfies the inhomogeneous transport equation
\begin{subequations} \label{eq:example:honf}
\begin{equation} \label{eq:example:honf:pde}
    \pderi_\theta \eta_3(\theta,t) + \pderi_t\eta_3(\theta,t) = -\alpha^\prime(\theta)y(t).
\end{equation}
Moreover, solving \eqref{eq:example:honf:fde:ausgeklammert} and \eqref{eq:example:honf:trafo:1}  for $\dot\eta_1(t)$ and $\dot\eta_2(t)$ yields the lumped dynamics
\begin{align}
    \dot \eta_1(t) &= -\frac{2k}{m}y(t), \\
    \dot \eta_2(t) &= \eta_1(t) -\frac{2k}{m}y(t).
\end{align}
Evaluation of $\eta_3(\theta,t)$ at both boundaries gives the \gls{bc}
\begin{align}
    \eta_3(0,t) &= \eta_2(t),
    \intertext{and the output equation}
    y(t) &= \eta_3(2,t).
\end{align}
\end{subequations}
The transformation from observability coordinates to the observer coordinates readily follows from \eqref{eq:example:honf:trafo} by
employing the definition  $\bar y(s,t)=y(t+s)$ of the observability coordinates.
This yields in particular
\begin{subequations}\label{eq:example:y_to_eta:final}
  \begin{align}
    \eta_1(t)&=\partial_\tau \bar{y}(0,t) +\partial_\tau {\bar{y}}(2,t) + \frac{1}{m}(\bar{y}(0,t) - \bar{y}(2,t))\notag \\ &\qquad +\frac{k}{m} \int_{0}^{2} \bar{y}(\btau,t) d\btau\\
    \eta_2(t)&=\bar{y}(0,t) +\bar{y}(2,t)+\int_0^2 \alpha'(\btau) \bar{y}(\btau,t)d\btau\\
    \eta_3(\tau,t) &=\bar y(2-\tau,t)+\int_\tau^2\bar y(\btau-\tau,t)\dd\alpha(\btau).
  \end{align}
\end{subequations}
Note that this transformation is well defined for $\bar y(\bullet,t)\in H^2([0,2])$.
\subsection{Transformation to original coordinates}
Eqs.\ \eqref{eq:example:param:x_by_y} and \eqref{eq:example:param:xi},   which allow the computation of the original state variables in $\systemstate(t)=(\pdestate(\bullet,t),\odestate(t))$ from the output trajectory,
depend on the restriction of $y$ to the interval $[t-1,t+1]$. However, the observability coordinates $\bar y$ correspond to
the restriction of $y$ to the interval $[t,t+2]$ (cf.\ Sec.~\ref{sec:hyperbolic:transform}). In order to compute $\systemstate(t)$ from $\bar y(\bullet,t)$, two alternatives have been discussed in
Sec.~\ref{sec:hyperbolic:transform}. Firstly, \eqref{eq:example:param:xi} can be evaluated at $t+1$ instead of $t$ yielding $\systemstate(t+1)$. Afterwards, the system equations are solved in negative time direction to obtain $\systemstate(t)$. Secondly, equation \eqref{eq:example:dde} is solved backward, providing the required restriction of $y$ to $[t-1,t+1]$. Afterwards, \eqref{eq:example:param:x_by_y} and \eqref{eq:example:param:xi} can be evaluated to obtain the original system state. In the following, the former alternative is pursued.
Combining \eqref{eq:example:y_to_eta:final} and \eqref{eq:example:param:xi}, and evaluating
\eqref{eq:example:param:x_by_y} immediately yields the system state at $t+1$:
\begin{subequations} \label{eq:example:trafo:eta_to_xi}
    \vspace{-1em}
  \begin{align}
    \xi_1(t+1) &= -\frac{m}{2k}\eta_1(t)+\frac{1}{2}\int_{0}^{2} \bar{y}(\btau,t) d\btau \\
    \xi_2(t+1) &= \frac12\left(\eta_2(t)-\int_0^2 \alpha'(\btau) \bar{y}(\btau,t)d\btau\right)\\
    x^{\mp}(z,t+1) &= \pm \bar y(1\mp(1-z),t).
   \end{align}
 \end{subequations}
 Observe that \eqref{eq:example:trafo:eta_to_xi} is well defined for $({\eta}_1(t),{\eta}_2(t),\bar{y}(\bullet,t))\in\reals^2\times\Lp([0,\hat\tau])$.
In order to obtain the system state $(\pdestate(t),\odestate(t))$ at time $t$ it remains to solve the modified system equations according to Section \ref{sec:hyperbolic:transform} in reverse time direction. For the present simple example, this can be  achieved rather easily. Firstly, consider the \gls{ode}
\begin{subequations}\label{eq:example:ode:reverse}
  \begin{align}
    \dot \xi_1(t) &= \xi_2(t), \\
    \dot \xi_2(t) &= -\frac{k}{m} \xi_1(t)-\frac{1}{m}\xi_2(t) +\frac1m y(t+1),
  \end{align}
\end{subequations}
which is obtained from \eqref{eq:example:system:ode:xi1}, \eqref{eq:example:system:ode:xi2}, \eqref{eq:example:bc0}, and \eqref{eq:example:param:x0}. Starting from $t+1$ this
equation is solved in reverse time direction to obtain $\odestate$ on the interval $[t,t+1]$.
Afterwards, the \gls{bc} \eqref{eq:example:bc0} is evaluated to compute $\transstatefwd(0,t+\tau)$ for $\tau\in [0,1]$. Finally, evaluating
\begin{align*}
  \transstatefwd(z,t)&=\transstatefwd(0,t+z)\\
  \transstatebwd(z,t)&=\transstatebwd(1,t+(1-z))=-\bar y(1-z,t)
\end{align*}
yields the distributed state $x(\bullet,t)$.

\section{Conclusion}\label{sec:conclusion}
This paper discusses the transformation to the \gls{honf} for linear hyperbolic
\glsentryshort{pde}--\glsentryshort{ode} systems with collocated  boundary
measurement.
To this end, the system state is parameterized in terms of the restriction of the output to a finite interval, which leads to a neutral \gls{fde} representation of the input--output dynamics.
Based on this, the \gls{honf} is derived,
in which distributed effects are captured by a transport
equation. The in-domain coupling between counter-propagating
states is handled via the kernel-based representation
introduced in Section~\ref{sec:hyperbolic}, where the state is expressed
through a combination of characteristic shifts and a
convolution-type integral operator. This operator captures
the effect of spatial coupling along the propagation paths.
The relation between the original system and the observer
coordinates is given by an invertible Volterra-type
transformation.
The proposed approach explicitly accounts for spatially distributed in-domain
coupling while still enabling a complete parameterization
of the system state through boundary measurements.
Despite this additional structural complexity, the \gls{honf} can be constructed 
systematically. The results are illustrated using a string--mass--spring
example, where the construction of the \gls{honf} is carried out explicitly.
Future work addresses observer design and extensions to
nonlinear boundary dynamics.



\bibliographystyle{IEEEtran}
\bibliography{bib/sources}

\end{document}